
\magnification 1200
\def\II{{\hbox{\tenrm I\kern-.19em{I}}}}

\font\abs=cmr9

\font\uit=cmu10

\def\uq2{U_q({\uit su}(2))}
\def\su2q{SU(2)_q}
\def\h1q{H(1)_q}
\def\e2q{E(2)_q}
\def\su2{{\uit su}(2)}
\def\r2{{\bf R}^2}
\def\j3{J_3}

\def\fraz#1#2{{\strut\displaystyle #1\over\displaystyle #2}}

\def\part#1#2{\fraz{\partial#1}{\partial#2}}

\def\ii#1{\item{$[#1]~$}}

\hsize= 15 truecm
\vsize= 22 truecm
\hoffset= 0.5 truecm
\voffset= 0 truecm

\baselineskip= 20 pt
\footline={\hss\tenrm\folio\hss} \pageno=1
\centerline
{\bf  INHOMOGENEOUS QUANTUM GROUPS }
\smallskip
\centerline
{\bf AS SYMMETRIES OF PHONONS.}

\bigskip\bigskip
\centerline{
{\it    F.Bonechi ${}^1$, E.Celeghini ${}^1$, R. Giachetti ${}^2$,
             E. Sorace ${}^1$ and M.Tarlini ${}^1$.}}
\bigskip
\baselineskip= 14 pt

   ${}^1$Dipartimento di Fisica, Universit\`a di Firenze and INFN--Firenze,
\footnote{}{\hskip -.85truecm E--mail: TARLINI@FI.INFN.IT\hfill
   DFF 152/12/91 Firenze}

   ${}^2$Dipartimento di Matematica, Universit\`a di Bologna and INFN--Firenze.
\bigskip
\bigskip
{\abs
\noindent{\bf Abstract.}
The quantum deformed $(1+1)$ Poincar\'e algebra is shown to be the
kinematical symmetry of the harmonic chain, whose spacing is given by
the deformation parameter. Phonons with their symmetries as well as
multiphonon processes are derived from the quantum group structure.
Inhomogeneous quantum groups are thus proposed as kinematical
invariance of discrete systems.}\hfil\break
\bigskip
\bigskip
\bigskip

Just from their early setting it has been observed the affinity
of quantum groups with finite difference calculus  [1] and the
presence of $q$--formulas in the realizations of these new algebras has
laid the ground for a theory of $q$--special functions [2].
It has also been shown that the quantum geometry associated
to the action of $q$--deformed groups on quantum spaces can be given by means
of $q$-difference operators [3].

Recently the quantization of inhomogeneous algebras of large physical interest
has been introduced [4,5,6] and a relevant point is that the
deformation parameter can acquire a dimension. This
possibility, realized by using a contraction procedure [4,5], results in a
meaningful tool for
the description of problems with a fundamental length scale.

In this letter we show that the $q$--deformed pseudoeuclidean algebra
$E_q(1,1)$ [4] arises quite naturally as kinematical symmetry of phonon
propagation in harmonic crystals, whose lattice spacing is interpreted as
the dimensional deformation parameter.
Even the coalgebra plays a fundamental physical
role as it establishes the rules for combining single phonons into composite
systems.

Let us consider the linear chain of equal masses lying at a distance $a$
from one another, with nearest neighbor harmonic interaction
and  subjected to an external restoring force.
The equations of motion are:
$$\ddot{z}_j(t)=\omega^2 \left(z_{j-1}(t)+z_{j+1}(t)-2z_j(t)\right)
-c^2 z_j(t)\ ,\eqno{(1)} $$
where $z_j(t)$ is the displacement of the $j$-th point $(j=0,1,\dots, N)$.
Periodic boundary conditions are assumed
and initial conditions $z_j(0),\ \dot{z}_j(0)$ must be specified.

We embed the ordinary system (1) into the PDE
$$ \left(\partial_t^2 + (2v/a)^2 \sin^2(-ia\partial_x/2)
+ m^2\right)\, z(x,t)=0\ ,
\eqno{(2)} $$
where $v=\omega a$, $m=c a$.
Letting $L=N a$, the periodic conditions are $z(0,t)=z(L,t)$ while
the Cauchy data for (2) consist in the assignment of smooth functions $z(x,0)$
and $\partial_t z(x,0)$. When $z(j a,0)=z_j(0),\ \partial_t z(j a,0)=
\dot{z}_j(0)$ for all $j\,$,  it is easy to see that the solutions
of (1) are directly obtained as $z_j(t)=z(j a,t)$ irrespectively of the
behaviour of the solutions of equation (2) in the points $x\not = ja$.

The continuum limit of (2) obviously reproduces the Klein--Gordon equation
in dimension $(1+1)$ with velocity $v$.
This constitutes a differential realization of the
Casimir of the $E(1,1)$ algebra, which is actually the kinematical symmetry
of the continuous system. Likewise, the differential operator in (2) gives
a realization of the Casimir invariant of the pseudoeuclidean quantum algebra
$E_q(1,1)$ [4] with $q=e^{i a}$ and real $a$ , which in its own right can be
considered the kinematical symmetry of the discrete system.

Indeed, in terms of the generators  $P_0,\; J,\; k,\; k^{-1}\;$
and of the unity $\II$ of the algebra,
the defining relations of $E_q(1,1)$ read
$$ \eqalign{
kP_0k^{-1}= & P_0,\quad ~~~~~~~ kJk^{-1}=J+aP_0,\quad ~~~~~~~ kk^{-1}=\II\ ,\cr
{} & {} ~~~~~~ JP_0-P_0J=(k-k^{-1})/(2a)\cr}\eqno{(3)}$$
The coproducts are
$$\eqalign{{} & \Delta(k)   = k \otimes k \cr
  {} & \Delta(P_0) = k^{-1/2}\otimes P_0 + P_0\otimes k^{1/2}\cr
  {} & \Delta(J)   = k^{-1/2}\otimes J  + J \otimes k^{1/2}\ ,\cr}
  \eqno{(4)}$$
while, for antipodes and counits we have
$$\gamma(J)=-J- (a/2)\; P_0,\quad ~~~\gamma(k)=k^{-1}\,
\quad ~~~\gamma(P_0)=-P_0,$$
$$\epsilon(J)=\epsilon(P_0)=0 \quad ~~~\epsilon(k)=1\ . $$
{}From equations (3) and (4) it appears that $k$ is a group-like element of
the Hopf algebra $E_q(1,1)$. If we write $k=\exp(iaP)$,
it is easy to see that $P$ is
a primitive element and it is determined up to an integer multiple of $2\pi/a$.
The Casimir operator is well defined in terms of $P$ and is given by:
$$C=P_0^2 - (2/a)^2 \sin^2(aP/2)\ .\eqno{(5)}$$

A realization of the relations (3) is obtained from [4], with minor changes,
yielding
$$P_0=(i/v)\ \partial_t\,, \quad ~~  k=\exp(a \partial_x)\,, \quad ~~
J=i(x/v)\partial_t-(vt/a)\sin(-ia\partial_x)\,.$$
Identifying $C=m^2/v^2$, the Casimir relation (5)
reproduces the Klein--Gordon equation on the lattice (2).

Here we analyze the case with $m=0$, which describes the phonons.
In the momentum representation, a realization of the $E_q(1,1)$
in terms of the diagonal $P$ and the position operator
$X=i \partial/\partial p$ is given by:
$$\eqalign{{} & P_0 = (2/a)\ \sin(ap/2),\quad\quad ~~~~~ 0\leq p<2\pi/a\ ,\cr
         {} &  J   = (1/a)\ \bigl\{\sin(ap/2),X\bigr\}_+\ ,\cr
         {} &  P   = p\ .}\eqno{(6)}$$
The limitation in the values of $p$ permits the reduction to
the first Brillouin zone, where $P_0$ has positive values.

The expression for the generator $J$ can be inverted in $X$:
$$ X=(1/2)\ \bigl\{P_0^{-1},J\bigr\}_+\ .\eqno{(7)}$$
The time derivative of $X$ is given by $\dot{X}=iv\ [P_0,X]$ and the
commutator, evaluated by the use of (7), gives the well known
group velocity of the phonons
$$\dot X = v_g= v\ \cos(aP/2)\ .$$

Let us show how the coproduct can be brought to bear to the study of the fusion
of phonons. It is well known that, when the symmetry is given by a Lie algebra,
the generators of the global symmetry of a composed system are obtained by
summing the generators of the symmetry of the elementary constituents. This
is related to the fact that
each generator $G$ of a Lie algebra is a primitive element,
{\it i.e.} $\Delta(G)=\II \otimes G + G\otimes \II \ $.
Then $G^{(1)}\equiv G\otimes \II \ $ acts
on the vector space of the first elementary system
and $G^{(2)}\equiv\II \otimes G$ on the second. The algebras generated by
$G^{(1)}$ and $G^{(2)}$ are both isomorphic to that generated by $G$ and since
$\Delta$ is a homomorphism of algebras, then $G^{(1)}+ G^{(2)}$
generates the same symmetry on the composed system.
In the quantum group context we can have non primitive generators, but
the very same considerations are still valid. The
coproducts (4) induce the global variables
$P_0 = e^{- i a P^{(1)}/2}\;P_0^{(2)} + P_0^{(1)}\;
e^{i aP^{(2)}/2}\ $,
$J = e^{- i a P^{(1)}/2}\;J^{(2)} + J^{(1)}\;e^{i a P^{(2)}/2}$
and $k=k^{(1)}k^{(2)}$ so that  $P = P^{(1)} + P^{(2)} + 2\pi n/a$,
where $n$ can be any integer value and will be chosen to keep $P$ in
the fixed Brillouin zone. The composition of the momenta
shows that the Umklapp process is implied by the quantum group symmetry.

In concrete, take
two differently polarized phonons with the same direction of propagation,
velocity parameters $v_1$ and $v_2$ and dispersion relations
(see {\it e.g.} [7]):
$$\Omega_1 = (2 v_1/a) \sin(aP^{(1)}/2), \quad ~~~
\Omega_2 =(2 v_2/a) \sin(aP^{(2)}/2)\ .$$
We then have
$P_0^{(1)}=\Omega_1/v_1$ and  $P_0^{(2)}=\Omega_2/v_2$.
The explicit coproduct of $P_0$ reads
$$\eqalign{
{} & P_0\ =\ e^{- i a P^{(1)}/2}\;(2/a) \sin(aP^{(2)}/2) +
     (2/a) \sin(aP^{(1)}/2)\;e^{i a P^{(2)}/2}\cr
{} & \phantom{P_0^{2}}\ =\ (2/a)\sin(a(P^{(1)}+P^{(2)})/2)\ .\cr}\eqno{(8)}$$

and the symmetry in the exchange of the two elementary components
together with the reality of the result is straightforward.
The energy conservation implies the existence of a global velocity $v$
such that, if $\Omega=|P_0| v=(2v/a)\sin(aP/2)$, then
$\Omega=\Omega_1+\Omega_2$ .

Contrary to energy and momentum, the global boost $J$ is not directly symmetric
and, to yield the correct statistics of
the composite system, it must be taken in the symmetrized form
$$J_s=\cos\left(aP^{(1)}/2\right)\, J^{(2)}+\cos\left(aP^{(2)}/2\right)\,
J^{(1)}\ .\eqno(9)$$
The generator $J_s$ is real and,
together with $P_0$ and $P$, still closes the $E_q(1,1)$ algebra.
Moreover from
equations (7) to (9) we get the
position operator of the composed phonon:
$$ X=\fraz 12 (X^{(1)}+X^{(2)}) + \fraz 12
\biggl\{\fraz{\sin(a(P^{(1)}-P^{(2)})/2)}{\sin(a(P^{(1)}+P^{(2)})/2)},
\fraz 12 (X^{(1)}-X^{(2)})\biggr\}_+\ ,\eqno{(10)}$$
which reproduces the Heisenberg algebra $\ [X,P]=i\ $ for the global variables.
Finally the group velocity of the composite system
$\dot{X}=i\ [\Omega,X]=v\ \cos(aP/2)\ $
appears formally identical to that of the elementary system, having performed
the Umklapp process.

We conclude by observing that the very same procedure can be applied to the
interaction of any number of phonons. Exactly as in the two phonon case it is
easily verified that energy and momentum come out in a directly symmetric
form. The global boost, first calculated from
the coproduct and then completely symmetrized,
close again the $E_q(1,1)$ algebra.
\bigskip
\bigskip

\centerline{{\bf References.}}

\bigskip

\ii 1 V. G. Drinfel'd, Proc. ICM Berkeley,
        (Providence, R.I., 1986, AMS) 798;\hfil\break
        N.Yu. Reshetikhin, L.A. Takhtadzhyan and L.D. Faddeev,
        Algebra and Analysis {\bf 1}, 193 (1990).
\smallskip

\ii 2 G.E. Andrews, Reg. Conf. Series in Math. n. 66 (Providence,
      R.I., 1986, AMS);\hfil\break
      L.L. Vaksman and L.I. Korogodskii, Soviet Math. Dokl. {\bf 39}, 173
      (1989);\hfil\break
      H.T. Koelink, {\it On quantum groups and $q$--special function},
      Ph.D. Thesis, (Leiden, Holland, 1991).
\smallskip

\ii 3 P.P. Kulish and E.V. Damashinsky, J. Phys. A: Math. Gen.,
      {\bf 23}, L415 (1990);\hfil\break
      M. Arik, Z. Phys. C: Particles and Fields, {\bf 51}, 627 (1991).
\smallskip

\ii 4 E. Celeghini, R. Giachetti, E. Sorace and M. Tarlini, J. Math. Phys.
      {\bf 31}, 2548 (1990); J. Math. Phys. {\bf 32}, 1155 (1991);
      J. Math. Phys. {\bf 32}, 1159 (1991);\hfil\break
      E. Celeghini, R. Giachetti, E. Sorace
      and M. Tarlini, ``{\it Contractions of quantum groups}'', Proceedings of
      the first semester on quantum groups, Eds. L.D. Faddeev
      and P.P. Kulish, Leningrad October 1990,
      Springer-Verlag, in press.
\smallskip

\ii 5 J. Lukierski, H. Ruegg, A. Nowicki and V.N. Tolstoy,
      Phys. Lett B {\bf 264}, 331 (1991).
\smallskip

\ii 6 V. Dobrev, {\it Canonical $q$--Deformation of non compact Lie (Super --)
      Algebras}, G\"ottingen University preprint, July 1991;\hfil\break
      J. Lukierski and A. Nowicki, {\it Quantum deformations of $D=4$
Poincar\'e
      and Weyl algebra from $q$--deformed $D=4$ conformal algebra}, University
      of Wroclaw  preprint , October 1991 ITP UWr 787/91.
\smallskip

\ii 7 N.W.Ashcroft, N.D.Mermin, {\it Solid State Physics}, (HRS International
      Editions, Philadelphia, PA, 1987).
\smallskip

\bye